\documentclass[graybox, envcountchap]{svmult}

\usepackage{mathptmx}        
\usepackage{amsmath}
\usepackage{amssymb}
\usepackage{color}
\usepackage{helvet}          
\usepackage{courier}         
\usepackage{dirtree}

\usepackage{makeidx}        
\usepackage{graphicx}        
\usepackage{subfig}

\usepackage{multicol}        
\usepackage[bottom]{footmisc}

\usepackage{hyperref}        
\hypersetup{colorlinks=true,urlcolor=blue}

\usepackage[misc]{ifsym}

\makeindex             

\def\beq{\begin{eqnarray}}
\def\eeq{\end{eqnarray}}

\usepackage{color}
\definecolor{darkgreen}{RGB}{0,120,0}

\definecolor{darkgreen}{RGB}{0,120,0}

\newcommand{\Mpch}{h^{-1}\mathrm{Mpc}}

\usepackage[square,sort,comma,numbers]{natbib}

\begin{document}


\title{Measuring $H_0$ with Spectroscopic Surveys}
\author{Mikhail M. Ivanov and Oliver H.~E. Philcox}
\institute{Mikhail M. Ivanov (\Letter) \at Institute for Advanced Study, Princeton \email{ivanov@ias.edu}
\and Oliver H.~E. Philcox (\Letter) \at Columbia University, New York \& Simons Foundation, New York \email{ohep2@cantab.ac.uk}}
%
%
\maketitle

\abstract{
Galaxy surveys map the three-dimensional distribution of matter in the Universe, encoding information about both the primordial cosmos and its subsequent evolution. By comparing the angular and physical scales of features in the galaxy distribution, we can compute the physical distance to the sample, and thus extract the Hubble parameter, $H_0$. In this chapter, we discuss how this is performed in practice, introducing two key ``standard rulers''. The first, the sound horizon at recombination, leads to baryon acoustic oscillations, and, by combining with external data from the CMB or Big Bang Nucleosynthesis, leads to a competitive $H_0$ constraint. Information can also be extracted from the physical scale of the horizon at matter-radiation equality; though somewhat less constraining, this depends on very different physics and is an important validation test of the physical model. We discuss how both such constraints can be derived (using ``template'' and ``full-shape'' methodologies), and present a number of recent constraints from the literature, some of which are comparable in precision to (and independent from) \textit{Planck}. Finally, we discuss future prospects for improving these constraints in the future.
%
%
}


\section{Introduction}\label{sec: intro}

Spectroscopic surveys map out the large-scale structure (hereafter LSS) of the Universe through the distribution of galaxies. By measuring the positions and redshifts of a wide variety of galaxies, we can build a three-dimensional map of the galaxy overdensity, which, on comparatively large-scales, traces the underlying density of dark matter \citep[e.g.,][]{1978IAUS...79..217P}. As such, spectroscopic surveys can be used to place constraints on physics operating in the early Universe (which set the initial conditions for matter clustering), and at late times (through the growth of structure). In this chapter, we will discuss how they can be used to place competitive constraints on the Hubble parameter, $H_0$.

Since the late 1970s, there have been a wealth of galaxy redshift surveys, with the current state-of-the-art public data (from the Sloan Digital Sky Survey \citep{SDSS:2011jap}) containing around one-million objects in the Northern sky. These are primarily analyzed through clustering statistics, the simplest of which are the power spectrum, $P(k)$, or two-point correlation function, $\xi(r)$, encoding the distribution of pairs of objects on some characteristic scale $r$ or $2\pi/k$ ($k$ denotes perturbations' wavenumber). 
In the coming decade the data volume will greatly increase, both due to ground-based (DESI, and, potentially, others such as MegaMapper) and space-based (SPHEREx, Euclid, Roman, and Rubin) missions \citep{DESI:2016fyo,Schlegel:2019eqc,Dore:2014cca,EUCLID:2011zbd,Tyson:2002mzo,Green:2012mj}. With increasing survey volume comes tighter constraints on cosmological parameters; as such, we may expect to yield sub-percent measurements of the Hubble parameter from a variety of probes in the near future.

As for the cosmic microwave background (CMB), spectroscopic $H_0$ constraints arise from the combination of observed and physical scales, through the canonical `standard ruler' mechanism. An important question, therefore, is which scales should be used. To be an accurate probe of $H_0$, a standard ruler must be: (a) precisely predicted form theory, and (b) well measured by experiment. In LSS, there are two candidates: the particle horizon scale at the epoch of matter-radiation equality (at $z_{\rm eq}\approx 3600$) \citep[e.g.,][]{Baxter:2020qlr} and the sound horizon at recombination (strictly, at baryon drag, $z_d\approx 1060$) \citep[e.g.,][]{Eisenstein:2006nj,SDSS:2005xqv} Each is straightforward to predict, and leaves detectable signatures in the observed galaxy clustering statistics, primarily (and fortuitously) concentrated in the linear regime. Whilst one could concoct other physical scales (such as that of Silk damping, or dark-energy-matter equality), the above are the simplest in practice in the context of galaxy surveys. 

Mathematically, the directly observed angular size of a feature, $\Delta\theta$ 
is related to the theoretically predicted physical size, $r$, via
\beq\label{eq:tfe}
    \Delta\theta_{\rm feature} \equiv \frac{r_{\rm feature}}{(1+z)D_A(z)},
\eeq
where $D_A(z)$ is the angular diameter distance to the galaxy survey at some redshift $z$. If the feature is primordial, $r_{\rm feature}$ does not depend on physics operating at late times, however $D_A(z)$ does, thus the measurement is not strictly `early' in nature. 

Let us use the following heuristic argument to understand where the 
sensitivity to $H_0$ comes from. 
Inserting the form of the angular diameter distance in Eq.~\eqref{eq:tfe} and assuming a flat Universe (for simplicity alone), we find
\beq\label{eq: ruler}
    H_0 = \frac{\Delta\theta_{\rm feature}}{r_{\rm feature}}\int_0^z\,\frac{dz'}{E(z')},
\eeq
where $E^2(z) = \Omega_{m,0}(1+z)^3+\Omega_{r,0}(1+z)^4+\Omega_\Lambda \equiv H^2(z)/H_0^2$.
Eq.~\eqref{eq: ruler} gives the rough idea on how the combination of a measured $r_{\rm feature}$ and an observed $\Delta\theta_{\rm feature}$ yields $H_0$ (given the cosmological expansion history); equivalently, the angular measurements translate (roughly) to constraints on $r_{\rm feature}H_0$. In the below, we will discuss our two possibilities for the standard ruler feature, including their physical origin and non-linear evolution, and present the various associated $H_0$ measurements, as well as their limitations and future prospects.\footnote{We caution that for the minimal $\Lambda$CDM model, and for  
probes of cosmological fluctuations such as the CMB and LSS, 
$\Omega_\Lambda, \Omega_m$ etc.
implicitely depend on $H_0$, so that Eq.~\eqref{eq: ruler}
does not actually \textit{define} the Hubble constant. } 

\section{Baryon Acoustic Oscillations}

\subsection{Theoretical Background}

Prior to recombination, the Universe is filled with a strongly interacting photon-baryon fluid, pressure-supported against. As with any fluid, this is filled with sound waves, on some characteristic scale $r_s$, seeded by early Universe inhomogeneities. As the Universe cools and free electrons combine with protons in recombination, the fluid decouples (at the baryon drag redshift $z\approx z_d\sim 1060$) and the photon-baryon oscillations cease to propagate, effectively `freezing' the sound waves at the sound-horizon scale, $r_s(z_d)\equiv r_d$. During the subsequent evolution of the Universe, this characteristic inhomogeneity is transferred to the dark matter distribution by gravitational interactions, yielding an excess of clustering at late-times.

Physically, these features (dubbed `baryon acoustic oscillations', hereafter BAO) source a peak in the two-point correlation function at $r\sim r_d$, or an oscillatory feature in the galaxy power spectrum at $k\sim 2\pi/r_d$. The precise sound horizon scale can be computed explicitly:
\beq\label{eq: BAO-scale}
    r_s(z) = \int_z^\infty dz'\frac{c_s(z)}{H(z')},
\eeq
with the sound speed $c_s \approx c\left[3+9/4\rho_b/\rho_\gamma\right]^{-1/2}$, where $\rho_{b,\gamma}$ are baryon and photon densities. All quantities in the above can be estimated precisely given cosmological parameters (such as those measured by \textit{Planck}), and are principally sensitive to the physical baryon density, $\omega_b$, since the photon density is strongly constrained from the CMB monopole. Assuming the \textit{Planck} $\Lambda$CDM cosmology, we find $r_d\sim 105\Mpch$, with an approximate cosmology dependence of \citep{Cuceu:2019for}
\beq
    r_d \approx \frac{55.154\,\mathrm{exp}\left[-72.3(\omega_{\nu,0}+0.0006)^2\right]}{\omega_{m,0}^{0.25351}\omega_{b,0}^{0.12807}}\mathrm{Mpc}\,,
\eeq
where $\omega_{m,0},\omega_{b,0},\omega_{\nu,0}$ denote the current physical 
densities of matter (i.e. dark matter + baryons), baryons, and neutrinos.

Whilst \eqref{eq: BAO-scale} accurately predicts the BAO scale at decoupling, various physical effects could alter its manifestation in the late-time galaxy correlators. Firstly, the amplitude of oscillatory features decays with time due to bulk flows, \textit{i.e.}\ the large-scale motion of matter in the post-recombination Universe. At late-times, the oscillatory part of the linear power spectrum, $P_{\rm L}^{\rm osc}$, (which contains the BAO wiggles), is suppressed according to 
\beq
\label{eq: bao-damping}
    & P_{\rm L}^{\rm osc}(k,z) \to e^{-k^2\Sigma^2(z)}P^{\rm osc}_{\rm L}(k,z), \\
    & \Sigma^2(z) = \frac{1}{6\pi^2}\int_0^{\Lambda}dp\,P_{\rm L}(p,z)[1-j_0(p r_d)
    +2j_2(p r_d)]\,,
\eeq
where $\Lambda\approx 0.1\,h\,\mathrm{Mpc}^{-1}$ is the IR cutoff scale, $j_\ell(x)$ are spherical Bessel
functions and 
$\Sigma(z)$ is BAO damping scales extracted via 
a systematic theoretical procedure called IR resummation~\cite{Senatore:2014via,Baldauf:2015xfa,Blas:2015qsi,Blas:2016sfa,Ivanov:2018gjr}
(also see~\cite{Crocce:2005xy,Crocce:2007dt} for early work on the subject).
When extracting the BAO signal from data, it is important to include such a suppression to avoid biased results. We note that this is \textit{not} included in simple parametrizations such as HaloFit.

Secondly, gravitational evolution modulates the clustering of matter at late times, causing various physical effects such as halo formation, which impact the galaxy power spectrum. This is important for $H_0$ determination, since non-linear physics could alter the position of peaks in the galaxy power spectrum, from which the absolute distance scale is extracted via \eqref{eq: ruler}. 
In practice, the effects of this are small, though not negligible \citep[e.g.,][]{Eisenstein:2006nj,Smith:2007gi}. 
Non-linear effects yield a (cosmology-dependent) $\sim 0.5\%$ shift in the BAO scale whose small amplitude is itself somewhat of an accident in $\Lambda$CDM~\cite{Sherwin:2012nh,Blas:2016sfa,Ivanov:2018gjr}.
In practice, one often marginalizes over non-linear contributions to the power spectrum, whilst assuming the BAO to be determined only by linear physics (plus damping) \citep[e.g.][]{BOSS:2016wmc}. This is sufficiently accurate for current surveys, but will not suffice in the future.

\subsection{Extracting the BAO Feature}
\subsubsection{Template Analyses}\label{subsec: scaling}
To infer $H_0$ from the BAO feature, it has become commonplace to use template (or scaling) analyses, based on the measurement of Alcock-Paczynski parameters \citep[e.g.,][]{Alcock:1979mp,BOSS:2016wmc}. This involves a theoretical templates for the BAO feature, such that the oscillatory part of the power spectrum (or equivalently correlation function), takes the form
\beq\label{eq: bao-template}
    P^{\rm obs}(k_\parallel,k_\perp) \propto P^{\rm template}(k_\parallel/\alpha_\parallel,k_\perp/\alpha_\perp),
\eeq
where we parametrize the three-dimensional space by $k_\parallel, k_\perp$ (relative to some line-of-sight direction), where the Alcock-Paczynski parameters are defined as
\beq
    \alpha_\parallel =  \frac{H^{\rm fid}(z)r_d^{\rm fid}}{H(z)r_d}\qquad \alpha_\perp = \frac{D_A(z)r^{\rm fid}_d}{D_A^{\rm fid}(z)r_d}
\eeq
These encode both the physical BAO scale, $r_d$, and the distance to the galaxy sample, via $D_A(z)$ (tangential) or $H(z)$ (radial, through velocities), and are normalized such that $\alpha_{\parallel,\perp}=1$ if the reference cosmology (used to define the templates) matches the truth. We caution that these parameters assume that the position of the BAO peak is not altered by non-linear physics.

In practice, one must additionally account for non-oscillatory effects contributing to the power spectrum, including galaxy bias, the fingers-of-God effect, and redshift-space distortions. Incorporating these effects (usually via some phenomenological framework, or with perturbation theory) gives rise to a full theoretical model for the galaxy power spectrum depending on a number of free parameters, including $\alpha_\parallel, \alpha_\perp$. Fitting this model to data then yields the best-fit parameters, themselves encoding $H(z)r_d$ and $D_A(z)/r_d$. Though these are usually said to be cosmology-independent, we caution that their estimation does require theoretical modeling to define the templates. If the templates are far from the truth, or new physics is at work that alters the BAO features, \eqref{eq: bao-template} breaks down, thus one cannot extract $H_0$-dependent quantities directly from the measured Alcock-Paczynski parameters. In practice, the former effect is small, assuming the background cosmology to be relatively well known.

The output of this procedure is measurements of the cosmological quantities $H(z)r_d$ and $D_A(z)/r_d$. Alone, these do not measure $H_0$, due to a perfect $H_0r_d$ degeneracy; practically, they are measuring only the angular size of the BAO feature. As mentioned above, one requires also the physical size. Combining these constraints with knowledge of $r_d$, within $\Lambda$CDM or some other model (such as early dark energy), one can break the degeneracy and measure $H_0$ directly. This is usually done by combining with the CMB \citep[e.g.,][]{BOSS:2016wmc}, or with Big Bang Nucleosynthesis probes \citep[e.g.,][]{Schoneberg:2019wmt}, both of which place strong constraints on $\omega_b$, and hence $r_d$. The resulting constraints on $H_0$ will be discussed in \S\ref{sec: meas}.

Whilst the above discussion has focused on measuring quantities from the observed galaxy power spectrum, it has become commonplace for survey analyses to include also data from the ``reconstructed'' power spectrum \citep{Eisenstein:2006nk}. In this instance, one first performs a catalog-level transformation of the data, effectively `undoing' the large-scale bulk flows which lead to smearing of the BAO feature. This can be performed robustly, and increases the signal-to-noise of oscillatory features by reducing the damping given in \eqref{eq: bao-damping}. The resulting BAO peak is well described by (weakly damped) linear theory (although we stress that the underlying physics relies on approximate cancellations between non-linear contributions), and can be used to enhance constraints on $H(z)r_d$ and $D_A(z)/r_d$. At low redshifts (where the damping is significant), this can bolster constraints by up to $\sim 40\%$.

\subsubsection{Full-Shape Analyses}\label{subsec: full-shape}
In the above formalism, the galaxy power spectrum is analyzed using a phenomenological model, effectively discarding any information besides the BAO feature. An alternative approach, which has found favor in recent years, is to instead model the entire power spectrum theoretically (not just the wiggles), from which a range of cosmological information can be extracted, including $H_0$ \citep[e.g.,][]{Ivanov:2019pdj,DAmico:2019fhj}. This is made possible via theoretical efforts such as the Effective Field Theory of Large Scale Structure (hereafter EFT, \citep[e.g.,][]{Baumann:2010tm,Carrasco:2012cv,Chudaykin:2020aoj,Ivanov:2022mrd,Cabass:2022avo}), which provides a first principles prediction for the galaxy power spectrum, $P_{gg}(k;\theta)$, based on perturbation theory and symmetry arguments, operating in the linear and mildly non-linear regime. This depends on a parameter vector $\theta$, which encompasses cosmology, galaxy bias, and other nuisance parameters. By fitting $P_{gg}(k;\theta)$ to the data directly, we obtain a posterior for $\theta$, and thus obtain constraints on the underlying cosmological parameters. Since the theory model depends on the cosmological initial conditions and subsequent evolution, it naturally picks up $H_0$ information through the BAO feature (and the equality scale, as discussed below), including both the physical size (from theory) and the angular size (distorted due to some choice of reference cosmology).

       Full-shape analyses are naturally dependent on the cosmological model; however, this is not a limitation, since one can construct predictions in any framework, be it $\Lambda$CDM or beyond, with extension parameters including early dark energy, modified gravity, and massive neutrinos. As far as $H_0$ is concerned, the approach is similar to a template fitting, but the templates themselves are dynamically varied, and thus depend on cosmological parameters such as $\Omega_m$ and (through the wiggle damping), $\omega_b$. Rather than predicting derived parameters such as $H(z)r_d)$, full-shape analyses can predict the underlying $H_0$ parameter (or some other combination of cosmological parameters, such as the CMB temperature, or angular sound horizon), since the physical forms of $D_A(z)$, $H(z)$, and $r_d$ are baked into the model. To extract $H_0$ with template-based approaches, external information on these functions must be supplied, thus BAO measurements are analyzed jointly with \textit{Planck}, or some other external data-set, which provide estimates of the $\Omega_m$ ($\omega_b$) parameters needed to determine $D_A$ and $H$ ($r_d$). In the full-shape case, both the parameters can be constrained from the data directly (through coordinate distortions and peak heights), though it is usually commonplace to supplement the survey with external $\omega_b$ information coming from the CMB or Big Bang Nucleosynthesis (BBN) constraints \citep[e.g.,][]{Aver:2015iza}. In the limit of a well-measured BAO feature, the precision of our output constraint on $H_0$ is fully limited by our knowledge of $\omega_b$, through the $r_d(\omega_b)$ relation: if this is not measured with full-shape approaches (which themselves give only comparatively weak constraints), we may not see large improvements in $H_0$ constraints if the external BBN information does not increase in precision.

As for template analyses, it is possible to consider the reconstructed power spectrum in the full-shape context. Once again, our logic is that reconstruction acts to sharpen the BAO oscillations by shifting information into the two-point function from the higher-point correlators, thus capturing $H_0$ information that would usually be found only in the bispectrum and above. Though several works have considered the full-shape modeling of the reconstructed power spectrum \citep{Hikage:2017tmm,Chen:2019lpf,Hikage:2019ihj,Ota:2021caz,Schmittfull:2017uhh}, it is highly non-trivial, and depends strongly on the particular flavor of reconstruction applied (quickly becoming unwieldy for multi-step reconstruction procedures). Given that the aim of reconstruction is to sharpen the BAO feature, and not to undo non-linear effects in the bulk flow (which it is known to perform poorly at, with non-linear formation being distorted by the necessary smoothing \citep{Schmittfull:2015mja}), it is thus appropriate to employ a hybrid approach, performing a template analysis of the reconstructed power spectrum and a full-shape analysis of the data, pre-reconstruction \citep[e.g.,][]{Philcox:2020vvt,Chen:2021wdi}. This proceeds by measuring Alcock-Paczynski parameters from the former (effectively acting as a data compression step), and performing a joint cosmological analysis of these with the measured power spectrum multipoles in a $\Lambda$CDM (or beyond) context, including the covariance between the two. This has been shown to bolster $H_0$ information from full-shape analyses by up to $\sim 40\%$ \citep{Philcox:2020vvt}, and avoids the difficulty in modeling broadband reconstruction artifacts.

              
 \section{Matter-Radiation Equality}\label{sec: equality}
\subsection{Theoretical Background}
 Around $z\approx 3600$, the dynamics of the Universe change from being dominated by radiation pressure, to being dominated by matter (both baryonic and dark). The characteristic redshift is straightforward to determine by setting equal the density of photons and matter, $\Omega_{m,0}(1+z)^3 = \Omega_{r,0}(1+z)^4$ (treating neutrinos as relativistic), where the radiation density is set by the current temperature of the CMB, as measured by FIRAS.\footnote{Note that physical moment of matter-radiation equality 
 is determined by the invariant matter-to-photon ratio
 and does not actually depend on the current CMB 
 temperature~\cite{Ivanov:2020mfr}.} 
 The horizon scale is set by the light horizon at this time, 
\beq\label{eq: eq-scale}
    r_{\rm eq} = \int_{z_{\rm eq}}^\infty dz'\frac{c}{H(z')},
\eeq
with the Fourier-space scale being approximately
\beq
	k_{\rm eq} = (2\Omega_m H_0^2z_{\rm eq})^{1/2}, \qquad z_{\rm eq} = 2.5 \times 10^4 \Omega_m h^2\Theta_{2.7}^{-4}
\eeq
\citep{Eisenstein:1997ik}, where $\Theta_{2.7}$ is the CMB temperature in $2.7$K units. This is a precisely known scale that acts as a second standard-ruler for $H_0$ analyses (dependent only on the physical matter density $\Omega_m h^2$).

Physically, the equality scale corresponds to a turn-over of the matter power spectrum at wavenumbers $k\sim k_{\rm eq}$, as described by the Meszaros equation. Whilst the turnover predominantly occurs on larger scales than can be robustly probed in many spectroscopic analyses (except for quasar samples, and SPHEREx, though it will be subject to significant cosmic variance and systematic penalities), its signature can be seen also on somewhat larger scales, due to a $\ln^2(k/k_{\rm eq})$ feature in the matter power spectrum at $k\geq k_{\rm eq}$. These scales are concentrated in the linear regime, which significantly simplifies modeling, especially given that the feature is spectrally smooth, unlike the easy-to-distinguish BAO oscillations. 

As before, we practically measure the ratio of the observed and theoretical equality scale: by the above arguments, this imparts a degeneracy $h\sim \Delta\theta_{\rm eq}k_{\rm eq}$, thus we in practice constrain the combination $\Omega_m h$, usually denoted $\Gamma$. The notion that galaxy surveys are sensitive to this scale arose significantly before the interest in the BAO feature, with measurements of $\Gamma$ being a primary target of the spectroscopic surveys of the 1990s and early 2000s \citep{Tegmark:1997rp,2dFGRS:2001csf}, though interest in its use as a BAO-independent probe of $H_0$ is comparatively recent.  

\subsection{Extracting Equality}
As for the BAO feature, one can proceed to measure the equality scale via template based approaches, by constructing some phenomenological model for the low-$k$ galaxy power spectrum, or from full-shape analyses. In this instance, most of the constraints have been derived using full-shape approaches, though there has been some work employed to connect the equality scale to effects parameters in the `ShapeFit' formalism \citep{Brieden:2021edu}. Though one could proceed to analyze only the galaxy power spectrum in the linear regime, it is usually preferred to utilize the whole EFTofLSS full-shape methodology to extract $H_0$ from the equality scale, since this (a) calibrates the galaxy bias parameters from quasi-linear scales, (b) naturally includes any non-linear contributions relevant at low-$k$, allowing their differentiation from the equality features in the transfer function. 

In practice, applying full-shape modeling to the galaxy power spectrum (and beyond) picks up information from both the equality and sound-horizon scales, since both enter the theory model $P_{gg}(k,\theta)$. To isolate the equality feature, one employ measures to remove the BAO features; it is not sufficient to simply employ a scale cut, since there is some overlap between the smooth equality feature and the first BAO wiggle. 
Noting that the BAO-derived $H_0$ information depends strongly on sound-horizon calibration through $\omega_b$ information, one option is to simply perform the analysis without an external prior on $\omega_b$ \citep{Philcox:2020xbv}. This is sufficient for current surveys (where the external information is much more precise than the $\omega_b$ measured internally, e.g., by BAO wiggle heights), but this will cease to be the case in the future, due to higher precision survey data. In this case, one may effectively marginalize over the oscillatory features, yielding an equality-only constraint on $H_0$ \citep{Farren:2021grl}. Notably that these measurements are naturally dependent on the choice of model used to analyze them (since one requires a prediction for $P_{gg}(k)$ in terms of $\Gamma \equiv \Omega_m h$), however, this is rarely a limitation since one can straightforwardly test any theoretical model of interest within the formalism, and, furthermore, there is very different dependence on the cosmological model to BAO analyses, e.g., the measurement would be largely unaffected by changes in sound-horizon physics. 

As described above, equality-based measurements are practically sensitive to $\Gamma \equiv \Omega_m h$, thus $H_0$ determination requires some additional constraint on $\Omega_m$ (and, to a lesser extent, $n_s$). This can be done within the survey itself, principally through coordinate distortions and redshift evolution, or externally. In the latter case, one can place a prior on $\Omega_m$ from uncalibrated BAO measurements or supernovae distance constraints (both of which measure the relative expansion rate, without $H_0$ calibration). We finally note that an analogous procedure can be used to constrain $H_0$ from photometric surveys or CMB lensing. The same does not hold for the BAO feature, since the projection integrals tend to wash out such features; indeed this allows straightforward baryon-independent $H_0$ constraints to be wrought.


\section{Measurements}\label{sec: meas}
\begin{figure}
    \centering
    \includegraphics[width=\textwidth]{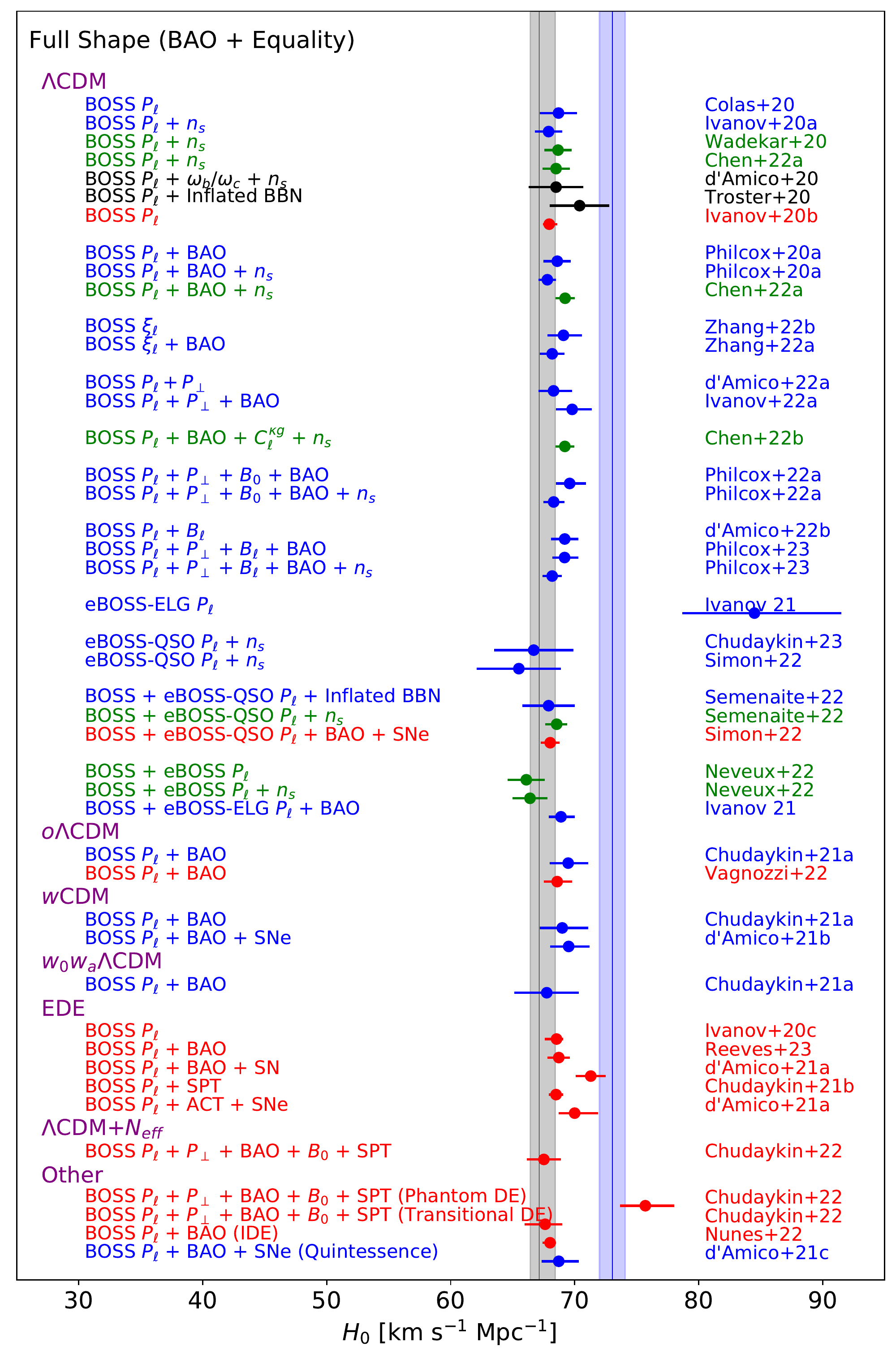}
    \caption{Summary plot of recent $H_0$ measurements from spectroscopic data, alongisde the \textit{Planck} CMB-only posterior (black) and SH0ES results (blue). In each case, we plot the $1\sigma$ bound on $H_0$ from the given analysis. Results shown in blue include a BBN prior on $\omega_b$, whilst those in green use an $\omega_b$ prior from \textit{Planck}. Measurements shown in red are combined with the full \textit{Planck} dataset. In each case we list the datasets used, with $P_\perp$ indicating real-space power spectrum proxies, $B_\ell$ indicating bispectrum multipoles, and $C_\ell^{\kappa g}$ indicating a galaxy-lensing cross spectrum. Entries denoted `+ SNe' add supernova $\Omega_m$ information from Pantheon or PantheonPlus, whilst `+ $n_s$' indicates use of the \textit{Planck} prior on $n_s$. This figure shows only full-shape results (which naturally include both BAO and equality information; Fig.\,\ref{fig: h02} shows the other measurements.}
    \label{fig: h01}
\end{figure}

\begin{figure}
    \centering
    \includegraphics[width=\textwidth]{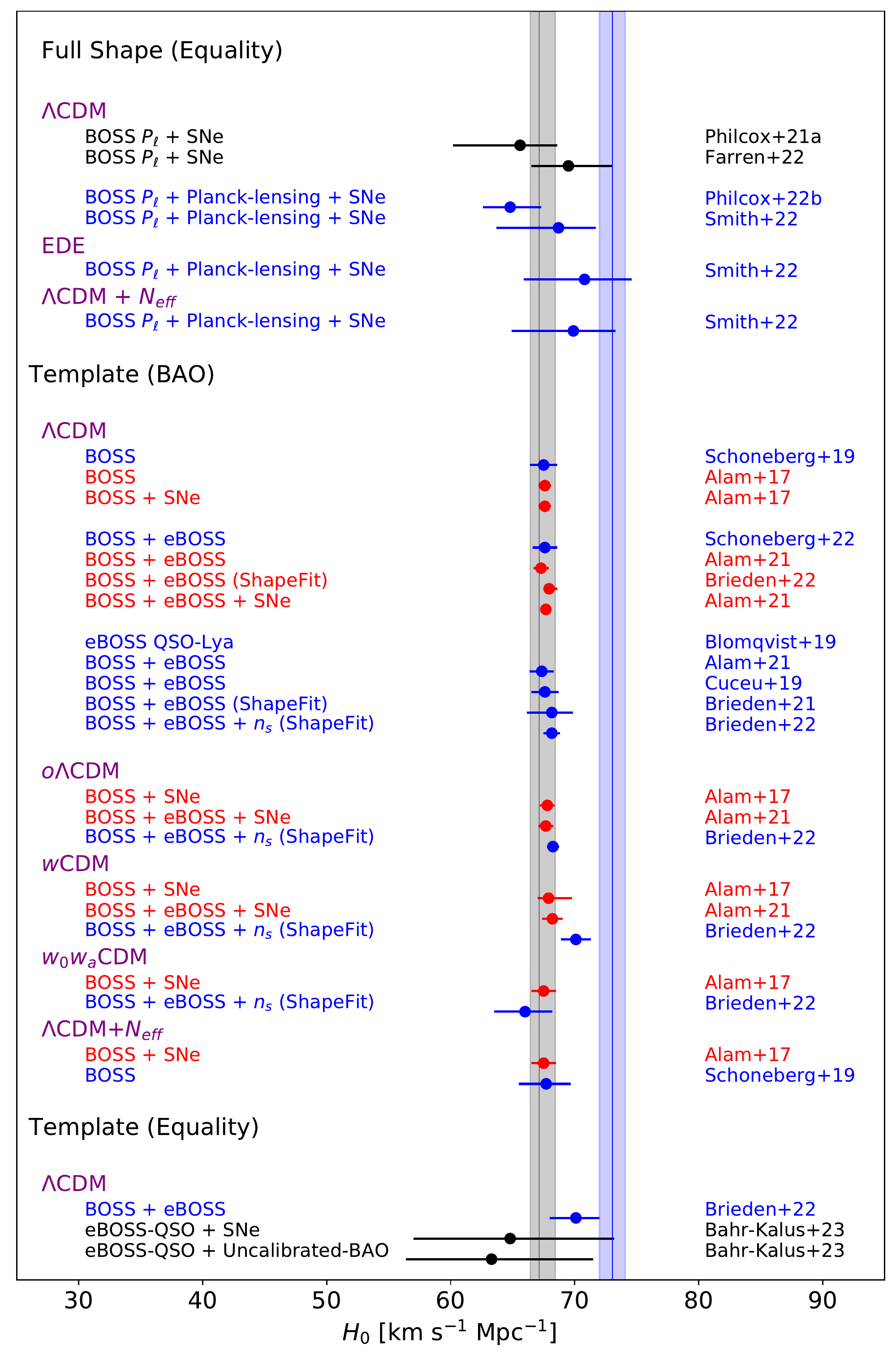}
    \caption{As Fig.\,\ref{fig: h01}, but displaying results from other types of measurement; full-shape equality measurements, template BAO constraints, and template equality results. The colors are as before. Results obtained using the `ShapeFit' compression scheme are indicated explicitly.}
    \label{fig: h02}
\end{figure}

Having discussed the physics behind spectroscopic $H_0$ constraints, we now present a number of recent constraints from both the BAO and equality scales. For tractability, we will specialize to analyses performed on BOSS data and beyond. A representative (though not necessarily complete) set of constraints are shown in Figs.\,\ref{fig: h01}\&\,\ref{fig: h02}, which indicate the plurality of work on the subject, particularly given that most results are derived from the same data-sets (BOSS and/or eBOSS clustering). In all cases, we assume a $\Lambda$CDM cosmology, unless otherwise specified. 

We first consider discuss sound-horizon-based constraints on $H_0$, as derived via template fitting. This is the preferred mode of analysis for the BOSS and eBOSS surveys, with the major data-product being a set of $H(z)r_d$ and $D_A(z)/r_d$ measurements at the effective redshifts of the relevant data sets. As discussed above, these can be combined with external `shape' information from the cosmic microwave background or Big Bang Nucleosynthesis (effectively fixing $\omega_b$), to break the $r_d$-degeneracy and extract $H_0$ directly. The official BOSS analysis (combining with \textit{Planck
}) finds $(67.6\pm0.5)\,\mathrm{km}\,\mathrm{s}^{-1}\mathrm{Mpc}^{-1}$  \citep{BOSS:2016wmc}; this is updated to $(67.6\pm0.4)\,\mathrm{km}\,\mathrm{s}^{-1}\mathrm{Mpc}^{-1}$  with the inclusion of eBOSS data \citep{eBOSS:2020yzd} (see also \citep{Blomqvist:2019rah,Cuceu:2019for}). Removing \textit{Planck} information leads to broader constraints (since \textit{Planck} tightly constrains $r_d$ and $\Omega_m$ as well as $H_0$ itself, through BAO features), with \citep{Schoneberg:2019wmt} finding $(67.5\pm1.1)\,\mathrm{km}\,\mathrm{s}^{-1}\mathrm{Mpc}^{-1}$, reducing to $(67.6\pm1.0)\,\mathrm{km}\,\mathrm{s}^{-1}\mathrm{Mpc}^{-1}$ with eBOSS \citep{Schoneberg:2022ggi} (see also \citep{Wang:2017yfu}).

Full-shape analyses make use of both sound-horizon and matter-radiation-equality standard rulers, which can yield tight constraints without explicit \textit{Planck} calibration (see also the \textit{ShapeFit} approach, which captures similar information in a template-based approach \citep{Brieden:2021edu,Brieden:2022lsd}, finding $(67.9_{-0.4}^{+0.7})\,\mathrm{km}\,\mathrm{s}^{-1}\mathrm{Mpc}^{-1}$ with BOSS, eBOSS and \textit{Planck}). A wide variety of full-shape $H_0$ constraints exist, predominantly differing in their choice of cosmological priors and supplementary datasets. For example, \citep{Colas:2019ret} find $(68.7\pm 1.5)\,\mathrm{km}\,\mathrm{s}^{-1}\mathrm{Mpc}^{-1}$ from the BOSS DR12 power spectrum with a \textit{Planck} prior on the baryon density (see also \citep{DAmico:2019fhj,Troster:2019ean}). Constraining the spectral slope to that measured by \textit{Planck} improves constraints by $\sim 40\%$, with consistent results found from both \textit{Planck} and BBN $\omega_b$ priors, and three major analysis teams \citep{Ivanov:2019pdj,Chen:2021wdi,Wadekar:2020hax,DAmico:2019fhj}. Combining with the full \textit{Planck} dataset yields $(68.0_{-0.5}^{+0.7})\,\mathrm{km}\,\mathrm{s}^{-1}\mathrm{Mpc}^{-1}$ \citep{Ivanov:2019hqk}).

The above results can be improved further by folding in information from the reconstructed power spectrum (which is standard practice in template analyses). As shown in \citep{Philcox:2020vvt}, this improves constraints by another $\sim 40\%$, with BOSS DR12 + BBN yielding $(67.8\pm0.7)\,\mathrm{km}\,\mathrm{s}^{-1}\mathrm{Mpc}^{-1}$ or $(68.6\pm1.1)\,\mathrm{km}\,\mathrm{s}^{-1}\mathrm{Mpc}^{-1}$ with and without $n_s$ priors (see also \citep{Chen:2021wdi,DAmico:2020ods}). A variety of extensions to these approaches have been performed, for example the inclusion of the configuration-space correlation function \citep{Chen:2021wdi,Zhang:2021yna}, a proxy for the real-space power spectrum 
\citep{Ivanov:2021fbu,DAmico:2021ymi} (also see ~\cite{Hamilton:2000du,Tegmark:2001jh,SDSS:2003tbn,Scoccimarro:2004tg}), lensing-galaxy cross correlations \citep{Chen:2022jzq}, the bispectrum monopole \citep{Philcox:2021kcw}, and the bispectrum multipoles \citep{Ivanov:2023qzb,DAmico:2022osl}. The most recent result, combining the BOSS power spectrum, reconstructed power spectrum, real-space power spectrum, and bispectrum multipoles, yields $(68.2\pm 0.8)\,\mathrm{km}\,\mathrm{s}^{-1}\mathrm{Mpc}^{-1}$ \citep{Ivanov:2023qzb}. This is consistent with the official BOSS results and those from \textit{Planck}. Similar results can be obtained using eBOSS data, both from the ELG \citep{Ivanov:2021zmi}, QSO  \citep{Ivanov:2021zmi,Semenaite:2021aen,Semenaite:2022unt,Chudaykin:2022nru,Simon:2022csv}, and combined samples \citep{Simon:2022csv,Neveux:2022tuk}, though their individual constraining power is weaker than that of BOSS.

Finally, we consider $H_0$ constraints which derive information only from the equality scale. Since this ruler suffers from increased cosmic variance and requires $\Omega_m$ calibration, the constraints are generally weaker than for the BAO, though can still be used to facilitate interesting tests of the cosmological model \citep{Farren:2021grl}. From template based analyses of the BOSS and eBOSS data with BBN information, we find $(70.1_{-2.1}^{+1.9})\,\mathrm{km}\,\mathrm{s}^{-1}\mathrm{Mpc}^{-1}$ using the `ShapeFit' formalism \citep{Brieden:2022heh}, whilst results from the eBOSS quasars yield the broad constraint $(64.8_{-7.8}^{+8.4})\,\mathrm{km}\,\mathrm{s}^{-1}\mathrm{Mpc}^{-1}$  \citep{Bahr-Kalus:2023ebd}. In full-shape analyses, one can isolate the equality information either by omitting the $\omega_b$ prior, or by explicit marginalization; combining BOSS with CMB lensing and `Pantheon+' supernovae, yields the final constraint $(64.8^{+2.2}_{-2.5})\,\mathrm{km}\,\mathrm{s}^{-1}\mathrm{Mpc}^{-1}$ , which is sound-horizon-independent but consistent with BAO constraints and \textit{Planck} (see also \citep{Philcox:2020xbv,Farren:2021grl,Baxter:2020qlr,Smith:2022iax}.

Whilst the above results have been derived in a $\Lambda$CDM framework, similar analyses are possible in any other cosmological model of choice. For the template analyses, one reinterprets the $H(z)r_d$ and $D_A(z)/r_d$ constraints with the updated model for growth and $r_d$, though we caution that this requires the desired cosmology to be relatively close to \textit{Planck}, since one assumed a $\Lambda$CDM model to generate the templates. In contrast, full-shape analyses can be extended by changing the underlying model used in computation of the linear power spectrum (and, if necessary, late-time evolution). A wide variety of alternative cosmologies have been considered, such as non-flat universes, varying dark energy, early dark energy, varying $N_{\rm eff}$, interacting dark energy, quintessence, \textit{et cetera} \citep{eBOSS:2020yzd,BOSS:2016wmc,Brieden:2022lsd}, full-shape BOSS [usually with Planck]: \citep{Chudaykin:2020ghx,Chudaykin:2022rnl,Vagnozzi:2020rcz,Kumar:2022vee,Reeves:2022aoi,DAmico:2020kxu,DAmico:2020tty,Ivanov:2020ril,Simon:2022adh}. In general, these give consistent constraints to those in $\Lambda$CDM but with broader errors, due to the additional degrees of freedom in the model.

\section{Summary \& Future Prospects}
Spectroscopic surveys can be utilized to provide stringent constraints on the Universe's structure, composition, and expansion rate. In this chapter, we have discussed their ability to measure the latter, through standard rulers provided by baryon acoustic oscillations and the transition from radiation- to matter-domination. In each case, one observes a characteristic angular size in the galaxy distribution (either directly or indirectly) which is compared to a theoretical computation, with the ratio giving an (inverse) absolute distance measurement, at known redshift. Historically, interest has focussed on the BAO signal analyzed by way of scaling templates; more recently, theoretical and computational advances have led to $H_0$ being directly extracted from the spectrum (through implicit standard ruler constraints) using full-shape methodologies. To date, these measurements have allowed for tight bounds on the Universe's expansion rate, with the BOSS survey yielding CMB-independent $\sim 1.6\%$ constraints on $H_0$ \citep[e.g.,][]{Ivanov:2023qzb}, which is of comparable width to the \textit{Planck} posterior.

\begin{figure}
    \centering
    \includegraphics[width=\textwidth]{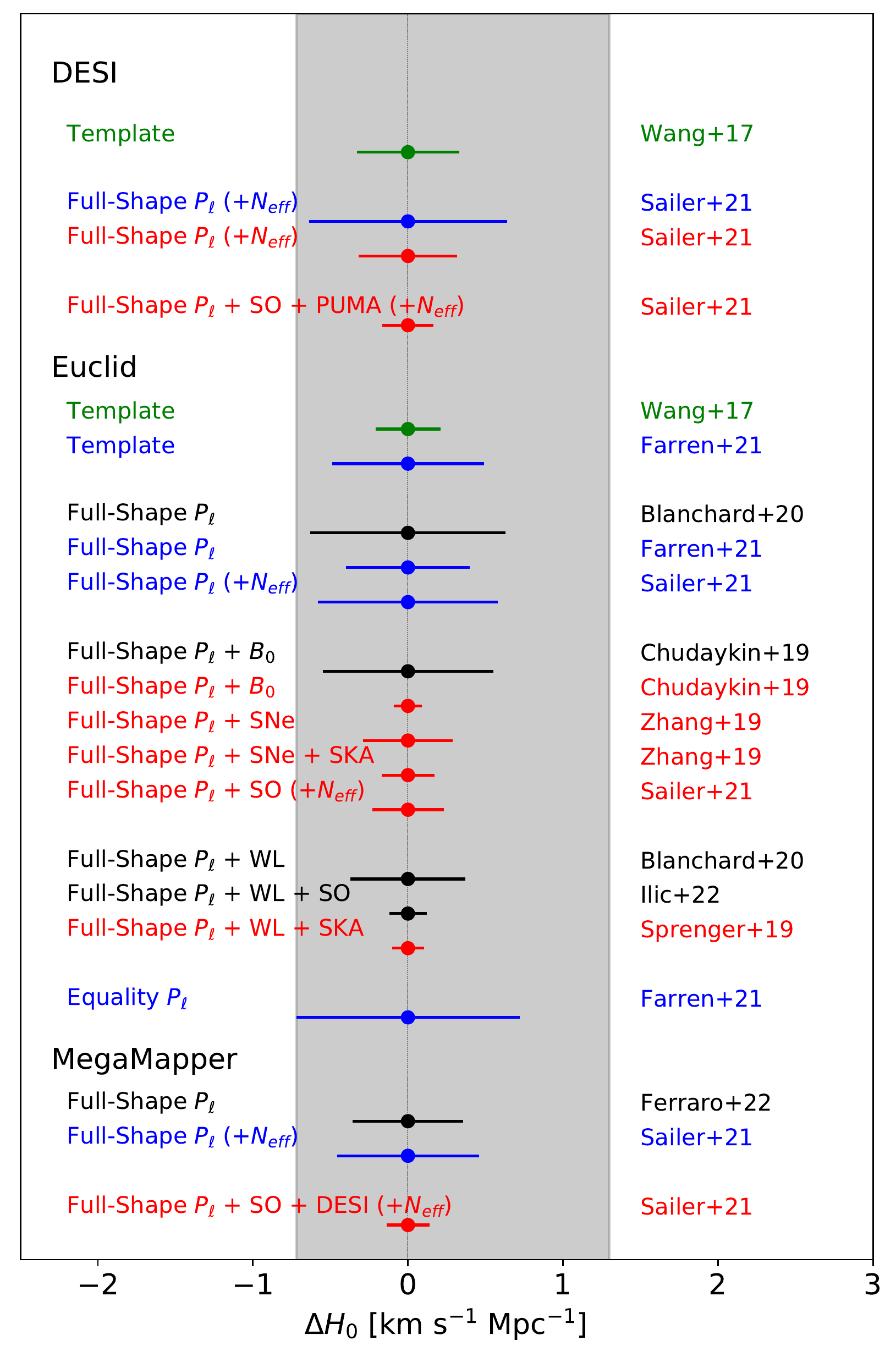}
    \caption{As Fig.\,\ref{fig: h01}, but displaying a collection of $H_0$ forecasts for future experiments from the literature. As before, constraints marked in blue include (current) BBN priors on $\omega_b$, those in green add \textit{Planck} priors on $\omega_b$, whilst those in red combine with the full \textit{Planck} posterior. Here, we group forecasts by instrument, with the captions denoting the forecasting set-up. `WL' indicates the inclusion of weak lensing information, whilst SO and SKA refer to the Simons Observatory and the Square Kilometer Array.}
    \label{fig: h03}
\end{figure}

It is interesting to discuss how such measurements will evolve in the future (see Fig.\,\ref{fig: h03} for a selection of representative forecasts). In the coming years, both DESI \citep{DESI:2016fyo} and Euclid \citep{EUCLID:2011zbd} will measure the redshifts of hundreds of millions of galaxies, providing precise mapping of the BAO parameters $H(z)r_d$ and $D_A(z)/r_d$ from low to intermediate redshifts; moreover, surveys such as SPHEREx \citep{Dore:2014cca} will add significant low-redshift measurements, and, more futuristically, proposed space missions including MegaMapper \citep{Schlegel:2019eqc} can yield a tight high-redshift extension. According to \citep{Wang:2017yfu}, five years of DESI data will lead to a tight BAO-derived $H_0$ constraint of $\sigma(H_0) = 0.33\,\mathrm{km}\,\mathrm{s}^{-1}\mathrm{Mpc}^{-1}$ via template methods, working at fixed $\omega_b$ (noting that the DESI white paper does not present $H_0$ constraints directly \citep{DESI:2016fyo}), whilst the same paper gives $\sigma(H_0) = 0.21\,\mathrm{km}\,\mathrm{s}^{-1}\mathrm{Mpc}^{-1}$ from \textit{Euclid}. As before, there are a plethora of forecasted constraints, with varying analysis choices: for example, the official Euclid forecast indicates a full-shape power spectrum constraint of $0.63\,\mathrm{km}\,\mathrm{s}^{-1}\mathrm{Mpc}^{-1}$ \citep{Euclid:2019clj} (without external priors), tightening to $0.37\,\mathrm{km}\,\mathrm{s}^{-1}\mathrm{Mpc}^{-1}$ in conjunction with weak lensing and photometric surveys, or $0.12\,\mathrm{km}\,\mathrm{s}^{-1}\mathrm{Mpc}^{-1}$, additionally adding in future CMB data \citep{Euclid:2021qvm} (see also \citep{Farren:2021grl, Chudaykin:2019ock,Zhang:2019dyq,Sprenger:2018tdb,Sailer:2021yzm}). For SPHEREx, no official  $H_0$ forecast has been made, but one can expect similar results to DESI and Euclid spectroscopy, but with most information coming from lower redshifts. In the future, surveys such as MegaMapper could yield $\sigma(H_0) = 0.36\,\mathrm{km}\,\mathrm{s}^{-1}\mathrm{Mpc}^{-1}$ alone, or three times tighter with CMB data, though the precise forecasts depend on which parameters are being varied (e.g., $N_{\rm eff}, M_\nu$) \citep{Ferraro:2022cmj,Sailer:2021yzm}. We note that many of these results are naturally reliant on the $\omega_b$ calibration (with BBN yielding a $\approx 1.6\%$ constraint \citep{Aver:2015iza}), thus, if this is tightened further (for example from better BBN measurements), one may expect a tighter increase still. 

From the equality scale, we may also expect significant improvements in the future. A forecast for Euclid found $\sigma(H_0) = 0.72\,\mathrm{km}\,\mathrm{s}^{-1}\mathrm{Mpc}^{-1}$ \citep{Farren:2021grl}, which is only some $40\%$ tighter than the BAO measurement (without reconstruction). We find that the two scales will be more balanced in terms of constraining power in the future, in the absence of tighter BBN calibration. As such, many of the above full-shape constraints are importantly impacted by equality-scale information. Finally, we note that a variety of other measurements will appear soon, both from additional galaxy surveys, such as Rubin and Roman \citep{Tyson:2002mzo,Green:2012mj}, as well as other probes including 21cm emission, line-intensity mapping, radio surveys, and beyond.

The coming measurements will open up a wealth of tests of the cosmological model. Firstly, BAO analyses allow one to map the $H(z)r_d$ and $D_A(z)/r_d$ derived parameters across a large swathe of redshifts, facilitating robust consistency tests of the low-redshift Universe; the same is true for full-shape derived constraints if we search for trends in the $H_0$ posteriors from different redshift samples. Secondly, the consistency of equality and BAO $H_0$ constraints has been previously shown to be a powerful diagnostic test of new physics in the early Universe, due to the different dependence on redshifts and recombination-scale physics. For now, the tests are weak, but with future surveys such as Euclid, one could potentially rule out many such models \citep{Farren:2021grl}. Furthermore, the two rulers are symbiotic; it is only by combining all sets of information available that we can obtain the strongest bounds on $H_0$, and thus physics new and old.  

\begin{acknowledgement}
We thank Eric Baxter, Giovanni Cabass, Gerrit Farren, Blake Sherwin, Marko Simonovic, and Matias Zaldarriaga, with whom many of the above $H_0$ constraints were derived. OHEP thanks Emirates for hosting a visit during which the majority of this draft was written. OHEP is a Junior Fellow of the Simons Society of Fellows and thanks the Simons Foundation for support.  
\end{acknowledgement}





\bibliographystyle{JHEP}
\bibliography{refs}

\providecommand{\href}[2]{#2}\begingroup\raggedright\begin{thebibliography}{10}

\bibitem{1978IAUS...79..217P}
P.~J.~E. {Peebles}, \emph{{Large Scale Clustering in the Universe}},  in
  \emph{Large Scale Structures in the Universe}, M.~S. {Longair} and
  J.~{Einasto}, eds., vol.~79, p.~217, Jan., 1978.

\bibitem{SDSS:2011jap}
{\scshape SDSS} collaboration, \emph{{SDSS-III: Massive Spectroscopic Surveys
  of the Distant Universe, the Milky Way Galaxy, and Extra-Solar Planetary
  Systems}}, \href{https://doi.org/10.1088/0004-6256/142/3/72}{\emph{Astron.
  J.} {\bfseries 142} (2011) 72}
  [\href{https://arxiv.org/abs/1101.1529}{{\ttfamily 1101.1529}}].

\bibitem{DESI:2016fyo}
{\scshape DESI} collaboration, \emph{{The DESI Experiment Part I:
  Science,Targeting, and Survey Design}},
  \href{https://arxiv.org/abs/1611.00036}{{\ttfamily 1611.00036}}.

\bibitem{Schlegel:2019eqc}
D.~J. Schlegel et~al., \emph{{Astro2020 APC White Paper: The MegaMapper: a z
  \ensuremath{>} 2 Spectroscopic Instrument for the Study of Inflation and Dark
  Energy}}, {\emph{Bull. Am. Astron. Soc.} {\bfseries 51} (2019) 229}
  [\href{https://arxiv.org/abs/1907.11171}{{\ttfamily 1907.11171}}].

\bibitem{Dore:2014cca}
O.~Dor\'e et~al., \emph{{Cosmology with the SPHEREX All-Sky Spectral Survey}},
  \href{https://arxiv.org/abs/1412.4872}{{\ttfamily 1412.4872}}.

\bibitem{EUCLID:2011zbd}
{\scshape EUCLID} collaboration, \emph{{Euclid Definition Study Report}},
  \href{https://arxiv.org/abs/1110.3193}{{\ttfamily 1110.3193}}.

\bibitem{Tyson:2002mzo}
{\scshape LSST} collaboration, \emph{{Large synoptic survey telescope:
  Overview}}, \href{https://doi.org/10.1117/12.456772}{\emph{Proc. SPIE Int.
  Soc. Opt. Eng.} {\bfseries 4836} (2002) 10}
  [\href{https://arxiv.org/abs/astro-ph/0302102}{{\ttfamily
  astro-ph/0302102}}].

\bibitem{Green:2012mj}
J.~Green et~al., \emph{{Wide-Field InfraRed Survey Telescope (WFIRST) Final
  Report}},  \href{https://arxiv.org/abs/1208.4012}{{\ttfamily 1208.4012}}.

\bibitem{Baxter:2020qlr}
E.~J. Baxter and B.~D. Sherwin, \emph{{Determining the Hubble Constant without
  the Sound Horizon Scale: Measurements from CMB Lensing}},
  \href{https://doi.org/10.1093/mnras/staa3706}{\emph{Mon. Not. Roy. Astron.
  Soc.} {\bfseries 501} (2021) 1823}
  [\href{https://arxiv.org/abs/2007.04007}{{\ttfamily 2007.04007}}].

\bibitem{Eisenstein:2006nj}
D.~J. Eisenstein, H.-j. Seo and M.~J. White, \emph{{On the Robustness of the
  Acoustic Scale in the Low-Redshift Clustering of Matter}},
  \href{https://doi.org/10.1086/518755}{\emph{Astrophys. J.} {\bfseries 664}
  (2007) 660} [\href{https://arxiv.org/abs/astro-ph/0604361}{{\ttfamily
  astro-ph/0604361}}].

\bibitem{SDSS:2005xqv}
{\scshape SDSS} collaboration, \emph{{Detection of the Baryon Acoustic Peak in
  the Large-Scale Correlation Function of SDSS Luminous Red Galaxies}},
  \href{https://doi.org/10.1086/466512}{\emph{Astrophys. J.} {\bfseries 633}
  (2005) 560} [\href{https://arxiv.org/abs/astro-ph/0501171}{{\ttfamily
  astro-ph/0501171}}].

\bibitem{Cuceu:2019for}
A.~Cuceu, J.~Farr, P.~Lemos and A.~Font-Ribera, \emph{{Baryon Acoustic
  Oscillations and the Hubble Constant: Past, Present and Future}},
  \href{https://doi.org/10.1088/1475-7516/2019/10/044}{\emph{JCAP} {\bfseries
  10} (2019) 044} [\href{https://arxiv.org/abs/1906.11628}{{\ttfamily
  1906.11628}}].

\bibitem{Senatore:2014via}
L.~Senatore and M.~Zaldarriaga, \emph{{The IR-resummed Effective Field Theory
  of Large Scale Structures}},
  \href{https://doi.org/10.1088/1475-7516/2015/02/013}{\emph{JCAP} {\bfseries
  02} (2015) 013} [\href{https://arxiv.org/abs/1404.5954}{{\ttfamily
  1404.5954}}].

\bibitem{Baldauf:2015xfa}
T.~Baldauf, M.~Mirbabayi, M.~Simonovi\'c and M.~Zaldarriaga, \emph{{Equivalence
  Principle and the Baryon Acoustic Peak}},
  \href{https://doi.org/10.1103/PhysRevD.92.043514}{\emph{Phys. Rev. D}
  {\bfseries 92} (2015) 043514}
  [\href{https://arxiv.org/abs/1504.04366}{{\ttfamily 1504.04366}}].

\bibitem{Blas:2015qsi}
D.~Blas, M.~Garny, M.~M. Ivanov and S.~Sibiryakov, \emph{{Time-Sliced
  Perturbation Theory for Large Scale Structure I: General Formalism}},
  \href{https://doi.org/10.1088/1475-7516/2016/07/052}{\emph{JCAP} {\bfseries
  07} (2016) 052} [\href{https://arxiv.org/abs/1512.05807}{{\ttfamily
  1512.05807}}].

\bibitem{Blas:2016sfa}
D.~Blas, M.~Garny, M.~M. Ivanov and S.~Sibiryakov, \emph{{Time-Sliced
  Perturbation Theory II: Baryon Acoustic Oscillations and Infrared
  Resummation}},
  \href{https://doi.org/10.1088/1475-7516/2016/07/028}{\emph{JCAP} {\bfseries
  07} (2016) 028} [\href{https://arxiv.org/abs/1605.02149}{{\ttfamily
  1605.02149}}].

\bibitem{Ivanov:2018gjr}
M.~M. Ivanov and S.~Sibiryakov, \emph{{Infrared Resummation for Biased Tracers
  in Redshift Space}},
  \href{https://doi.org/10.1088/1475-7516/2018/07/053}{\emph{JCAP} {\bfseries
  07} (2018) 053} [\href{https://arxiv.org/abs/1804.05080}{{\ttfamily
  1804.05080}}].

\bibitem{Crocce:2005xy}
M.~Crocce and R.~Scoccimarro, \emph{{Renormalized cosmological perturbation
  theory}}, \href{https://doi.org/10.1103/PhysRevD.73.063519}{\emph{Phys. Rev.
  D} {\bfseries 73} (2006) 063519}
  [\href{https://arxiv.org/abs/astro-ph/0509418}{{\ttfamily
  astro-ph/0509418}}].

\bibitem{Crocce:2007dt}
M.~Crocce and R.~Scoccimarro, \emph{{Nonlinear Evolution of Baryon Acoustic
  Oscillations}}, \href{https://doi.org/10.1103/PhysRevD.77.023533}{\emph{Phys.
  Rev. D} {\bfseries 77} (2008) 023533}
  [\href{https://arxiv.org/abs/0704.2783}{{\ttfamily 0704.2783}}].

\bibitem{Smith:2007gi}
R.~E. Smith, R.~Scoccimarro and R.~K. Sheth, \emph{{Eppur Si Muove: On The
  Motion of the Acoustic Peak in the Correlation Function}},
  \href{https://doi.org/10.1103/PhysRevD.77.043525}{\emph{Phys. Rev. D}
  {\bfseries 77} (2008) 043525}
  [\href{https://arxiv.org/abs/astro-ph/0703620}{{\ttfamily
  astro-ph/0703620}}].

\bibitem{Sherwin:2012nh}
B.~D. Sherwin and M.~Zaldarriaga, \emph{{The Shift of the Baryon Acoustic
  Oscillation Scale: A Simple Physical Picture}},
  \href{https://doi.org/10.1103/PhysRevD.85.103523}{\emph{Phys. Rev. D}
  {\bfseries 85} (2012) 103523}
  [\href{https://arxiv.org/abs/1202.3998}{{\ttfamily 1202.3998}}].

\bibitem{BOSS:2016wmc}
{\scshape BOSS} collaboration, \emph{{The clustering of galaxies in the
  completed SDSS-III Baryon Oscillation Spectroscopic Survey: cosmological
  analysis of the DR12 galaxy sample}},
  \href{https://doi.org/10.1093/mnras/stx721}{\emph{Mon. Not. Roy. Astron.
  Soc.} {\bfseries 470} (2017) 2617}
  [\href{https://arxiv.org/abs/1607.03155}{{\ttfamily 1607.03155}}].

\bibitem{Alcock:1979mp}
C.~Alcock and B.~Paczynski, \emph{{An evolution free test for non-zero
  cosmological constant}},
  \href{https://doi.org/10.1038/281358a0}{\emph{Nature} {\bfseries 281} (1979)
  358}.

\bibitem{Schoneberg:2019wmt}
N.~Sch\"oneberg, J.~Lesgourgues and D.~C. Hooper, \emph{{The BAO+BBN take on
  the Hubble tension}},
  \href{https://doi.org/10.1088/1475-7516/2019/10/029}{\emph{JCAP} {\bfseries
  10} (2019) 029} [\href{https://arxiv.org/abs/1907.11594}{{\ttfamily
  1907.11594}}].

\bibitem{Eisenstein:2006nk}
D.~J. Eisenstein, H.-j. Seo, E.~Sirko and D.~Spergel, \emph{{Improving
  Cosmological Distance Measurements by Reconstruction of the Baryon Acoustic
  Peak}}, \href{https://doi.org/10.1086/518712}{\emph{Astrophys. J.} {\bfseries
  664} (2007) 675} [\href{https://arxiv.org/abs/astro-ph/0604362}{{\ttfamily
  astro-ph/0604362}}].

\bibitem{Ivanov:2019pdj}
M.~M. Ivanov, M.~Simonovi\'c and M.~Zaldarriaga, \emph{{Cosmological Parameters
  from the BOSS Galaxy Power Spectrum}},
  \href{https://doi.org/10.1088/1475-7516/2020/05/042}{\emph{JCAP} {\bfseries
  05} (2020) 042} [\href{https://arxiv.org/abs/1909.05277}{{\ttfamily
  1909.05277}}].

\bibitem{DAmico:2019fhj}
G.~D'Amico, J.~Gleyzes, N.~Kokron, K.~Markovic, L.~Senatore, P.~Zhang et~al.,
  \emph{{The Cosmological Analysis of the SDSS/BOSS data from the Effective
  Field Theory of Large-Scale Structure}},
  \href{https://doi.org/10.1088/1475-7516/2020/05/005}{\emph{JCAP} {\bfseries
  05} (2020) 005} [\href{https://arxiv.org/abs/1909.05271}{{\ttfamily
  1909.05271}}].

\bibitem{Baumann:2010tm}
D.~Baumann, A.~Nicolis, L.~Senatore and M.~Zaldarriaga, \emph{{Cosmological
  Non-Linearities as an Effective Fluid}},
  \href{https://doi.org/10.1088/1475-7516/2012/07/051}{\emph{JCAP} {\bfseries
  07} (2012) 051} [\href{https://arxiv.org/abs/1004.2488}{{\ttfamily
  1004.2488}}].

\bibitem{Carrasco:2012cv}
J.~J.~M. Carrasco, M.~P. Hertzberg and L.~Senatore, \emph{{The Effective Field
  Theory of Cosmological Large Scale Structures}},
  \href{https://doi.org/10.1007/JHEP09(2012)082}{\emph{JHEP} {\bfseries 09}
  (2012) 082} [\href{https://arxiv.org/abs/1206.2926}{{\ttfamily 1206.2926}}].

\bibitem{Chudaykin:2020aoj}
A.~Chudaykin, M.~M. Ivanov, O.~H.~E. Philcox and M.~Simonovi\'c,
  \emph{{Nonlinear perturbation theory extension of the Boltzmann code CLASS}},
  \href{https://doi.org/10.1103/PhysRevD.102.063533}{\emph{Phys. Rev. D}
  {\bfseries 102} (2020) 063533}
  [\href{https://arxiv.org/abs/2004.10607}{{\ttfamily 2004.10607}}].

\bibitem{Ivanov:2022mrd}
M.~M. Ivanov, \emph{{Effective Field Theory for Large Scale Structure}},
  \href{https://arxiv.org/abs/2212.08488}{{\ttfamily 2212.08488}}.

\bibitem{Cabass:2022avo}
G.~Cabass, M.~M. Ivanov, M.~Lewandowski, M.~Mirbabayi and M.~Simonovi\'c,
  \emph{{Snowmass White Paper: Effective Field Theories in Cosmology}},  in
  \emph{{2022 Snowmass Summer Study}}, 3, 2022,
  \href{https://arxiv.org/abs/2203.08232}{{\ttfamily 2203.08232}}.

\bibitem{Aver:2015iza}
E.~Aver, K.~A. Olive and E.~D. Skillman, \emph{{The effects of He I
  \ensuremath{\lambda}10830 on helium abundance determinations}},
  \href{https://doi.org/10.1088/1475-7516/2015/07/011}{\emph{JCAP} {\bfseries
  07} (2015) 011} [\href{https://arxiv.org/abs/1503.08146}{{\ttfamily
  1503.08146}}].

\bibitem{Hikage:2017tmm}
C.~Hikage, K.~Koyama and A.~Heavens, \emph{{Perturbation theory for BAO
  reconstructed fields: One-loop results in the real-space matter density
  field}}, \href{https://doi.org/10.1103/PhysRevD.96.043513}{\emph{Phys. Rev.
  D} {\bfseries 96} (2017) 043513}
  [\href{https://arxiv.org/abs/1703.07878}{{\ttfamily 1703.07878}}].

\bibitem{Chen:2019lpf}
S.-F. Chen, Z.~Vlah and M.~White, \emph{{The reconstructed power spectrum in
  the Zeldovich approximation}},
  \href{https://doi.org/10.1088/1475-7516/2019/09/017}{\emph{JCAP} {\bfseries
  09} (2019) 017} [\href{https://arxiv.org/abs/1907.00043}{{\ttfamily
  1907.00043}}].

\bibitem{Hikage:2019ihj}
C.~Hikage, K.~Koyama and R.~Takahashi, \emph{{Perturbation theory for the
  redshift-space matter power spectra after reconstruction}},
  \href{https://doi.org/10.1103/PhysRevD.101.043510}{\emph{Phys. Rev. D}
  {\bfseries 101} (2020) 043510}
  [\href{https://arxiv.org/abs/1911.06461}{{\ttfamily 1911.06461}}].

\bibitem{Ota:2021caz}
A.~Ota, H.-J. Seo, S.~Saito and F.~Beutler, \emph{{Modeling iterative
  reconstruction and displacement field in the large scale structure}},
  \href{https://doi.org/10.1103/PhysRevD.104.123508}{\emph{Phys. Rev. D}
  {\bfseries 104} (2021) 123508}
  [\href{https://arxiv.org/abs/2106.00146}{{\ttfamily 2106.00146}}].

\bibitem{Schmittfull:2017uhh}
M.~Schmittfull, T.~Baldauf and M.~Zaldarriaga, \emph{{Iterative initial
  condition reconstruction}},
  \href{https://doi.org/10.1103/PhysRevD.96.023505}{\emph{Phys. Rev. D}
  {\bfseries 96} (2017) 023505}
  [\href{https://arxiv.org/abs/1704.06634}{{\ttfamily 1704.06634}}].

\bibitem{Schmittfull:2015mja}
M.~Schmittfull, Y.~Feng, F.~Beutler, B.~Sherwin and M.~Y. Chu, \emph{{Eulerian
  BAO Reconstructions and N-Point Statistics}},
  \href{https://doi.org/10.1103/PhysRevD.92.123522}{\emph{Phys. Rev. D}
  {\bfseries 92} (2015) 123522}
  [\href{https://arxiv.org/abs/1508.06972}{{\ttfamily 1508.06972}}].

\bibitem{Philcox:2020vvt}
O.~H.~E. Philcox, M.~M. Ivanov, M.~Simonovi\'c and M.~Zaldarriaga,
  \emph{{Combining Full-Shape and BAO Analyses of Galaxy Power Spectra: A
  1.6\textbackslash{}\% CMB-independent constraint on H$_0$}},
  \href{https://doi.org/10.1088/1475-7516/2020/05/032}{\emph{JCAP} {\bfseries
  05} (2020) 032} [\href{https://arxiv.org/abs/2002.04035}{{\ttfamily
  2002.04035}}].

\bibitem{Chen:2021wdi}
S.-F. Chen, Z.~Vlah and M.~White, \emph{{A new analysis of galaxy 2-point
  functions in the BOSS survey, including full-shape information and
  post-reconstruction BAO}},
  \href{https://doi.org/10.1088/1475-7516/2022/02/008}{\emph{JCAP} {\bfseries
  02} (2022) 008} [\href{https://arxiv.org/abs/2110.05530}{{\ttfamily
  2110.05530}}].

\bibitem{Ivanov:2020mfr}
M.~M. Ivanov, Y.~Ali-Ha\"\i{}moud and J.~Lesgourgues, \emph{{H0 tension or T0
  tension?}}, \href{https://doi.org/10.1103/PhysRevD.102.063515}{\emph{Phys.
  Rev. D} {\bfseries 102} (2020) 063515}
  [\href{https://arxiv.org/abs/2005.10656}{{\ttfamily 2005.10656}}].

\bibitem{Eisenstein:1997ik}
D.~J. Eisenstein and W.~Hu, \emph{{Baryonic features in the matter transfer
  function}}, \href{https://doi.org/10.1086/305424}{\emph{Astrophys. J.}
  {\bfseries 496} (1998) 605}
  [\href{https://arxiv.org/abs/astro-ph/9709112}{{\ttfamily
  astro-ph/9709112}}].

\bibitem{Tegmark:1997rp}
M.~Tegmark, \emph{{Measuring cosmological parameters with galaxy surveys}},
  \href{https://doi.org/10.1103/PhysRevLett.79.3806}{\emph{Phys. Rev. Lett.}
  {\bfseries 79} (1997) 3806}
  [\href{https://arxiv.org/abs/astro-ph/9706198}{{\ttfamily
  astro-ph/9706198}}].

\bibitem{2dFGRS:2001csf}
{\scshape 2dFGRS} collaboration, \emph{{The 2dF Galaxy Redshift Survey: The
  Power spectrum and the matter content of the Universe}},
  \href{https://doi.org/10.1046/j.1365-8711.2001.04827.x}{\emph{Mon. Not. Roy.
  Astron. Soc.} {\bfseries 327} (2001) 1297}
  [\href{https://arxiv.org/abs/astro-ph/0105252}{{\ttfamily
  astro-ph/0105252}}].

\bibitem{Brieden:2021edu}
S.~Brieden, H.~Gil-Mar\'\i{}n and L.~Verde, \emph{{ShapeFit: extracting the
  power spectrum shape information in galaxy surveys beyond BAO and RSD}},
  \href{https://doi.org/10.1088/1475-7516/2021/12/054}{\emph{JCAP} {\bfseries
  12} (2021) 054} [\href{https://arxiv.org/abs/2106.07641}{{\ttfamily
  2106.07641}}].

\bibitem{Philcox:2020xbv}
O.~H.~E. Philcox, B.~D. Sherwin, G.~S. Farren and E.~J. Baxter,
  \emph{{Determining the Hubble Constant without the Sound Horizon:
  Measurements from Galaxy Surveys}},
  \href{https://doi.org/10.1103/PhysRevD.103.023538}{\emph{Phys. Rev. D}
  {\bfseries 103} (2021) 023538}
  [\href{https://arxiv.org/abs/2008.08084}{{\ttfamily 2008.08084}}].

\bibitem{Farren:2021grl}
G.~S. Farren, O.~H.~E. Philcox and B.~D. Sherwin, \emph{{Determining the Hubble
  constant without the sound horizon: Perspectives with future galaxy
  surveys}}, \href{https://doi.org/10.1103/PhysRevD.105.063503}{\emph{Phys.
  Rev. D} {\bfseries 105} (2022) 063503}
  [\href{https://arxiv.org/abs/2112.10749}{{\ttfamily 2112.10749}}].

\bibitem{eBOSS:2020yzd}
{\scshape eBOSS} collaboration, \emph{{Completed SDSS-IV extended Baryon
  Oscillation Spectroscopic Survey: Cosmological implications from two decades
  of spectroscopic surveys at the Apache Point Observatory}},
  \href{https://doi.org/10.1103/PhysRevD.103.083533}{\emph{Phys. Rev. D}
  {\bfseries 103} (2021) 083533}
  [\href{https://arxiv.org/abs/2007.08991}{{\ttfamily 2007.08991}}].

\bibitem{Blomqvist:2019rah}
M.~Blomqvist et~al., \emph{{Baryon acoustic oscillations from the
  cross-correlation of Ly$\alpha$ absorption and quasars in eBOSS DR14}},
  \href{https://doi.org/10.1051/0004-6361/201935641}{\emph{Astron. Astrophys.}
  {\bfseries 629} (2019) A86}
  [\href{https://arxiv.org/abs/1904.03430}{{\ttfamily 1904.03430}}].

\bibitem{Schoneberg:2022ggi}
N.~Sch\"oneberg, L.~Verde, H.~Gil-Mar\'\i{}n and S.~Brieden, \emph{{BAO+BBN
  revisited -- growing the Hubble tension with a 0.7 km/s/Mpc constraint}},
  \href{https://doi.org/10.1088/1475-7516/2022/11/039}{\emph{JCAP} {\bfseries
  11} (2022) 039} [\href{https://arxiv.org/abs/2209.14330}{{\ttfamily
  2209.14330}}].

\bibitem{Wang:2017yfu}
Y.~Wang, L.~Xu and G.-B. Zhao, \emph{{A measurement of the Hubble constant
  using galaxy redshift surveys}},
  \href{https://doi.org/10.3847/1538-4357/aa8f48}{\emph{Astrophys. J.}
  {\bfseries 849} (2017) 84}
  [\href{https://arxiv.org/abs/1706.09149}{{\ttfamily 1706.09149}}].

\bibitem{Brieden:2022lsd}
S.~Brieden, H.~Gil-Mar\'\i{}n and L.~Verde, \emph{{Model-agnostic
  interpretation of 10 billion years of cosmic evolution traced by BOSS and
  eBOSS data}},
  \href{https://doi.org/10.1088/1475-7516/2022/08/024}{\emph{JCAP} {\bfseries
  08} (2022) 024} [\href{https://arxiv.org/abs/2204.11868}{{\ttfamily
  2204.11868}}].

\bibitem{Colas:2019ret}
T.~Colas, G.~D'amico, L.~Senatore, P.~Zhang and F.~Beutler, \emph{{Efficient
  Cosmological Analysis of the SDSS/BOSS data from the Effective Field Theory
  of Large-Scale Structure}},
  \href{https://doi.org/10.1088/1475-7516/2020/06/001}{\emph{JCAP} {\bfseries
  06} (2020) 001} [\href{https://arxiv.org/abs/1909.07951}{{\ttfamily
  1909.07951}}].

\bibitem{Troster:2019ean}
T.~Tr\"oster et~al., \emph{{Cosmology from large-scale structure: Constraining
  $\Lambda$CDM with BOSS}},
  \href{https://doi.org/10.1051/0004-6361/201936772}{\emph{Astron. Astrophys.}
  {\bfseries 633} (2020) L10}
  [\href{https://arxiv.org/abs/1909.11006}{{\ttfamily 1909.11006}}].

\bibitem{Wadekar:2020hax}
D.~Wadekar, M.~M. Ivanov and R.~Scoccimarro, \emph{{Cosmological constraints
  from BOSS with analytic covariance matrices}},
  \href{https://doi.org/10.1103/PhysRevD.102.123521}{\emph{Phys. Rev. D}
  {\bfseries 102} (2020) 123521}
  [\href{https://arxiv.org/abs/2009.00622}{{\ttfamily 2009.00622}}].

\bibitem{Ivanov:2019hqk}
M.~M. Ivanov, M.~Simonovi\'c and M.~Zaldarriaga, \emph{{Cosmological Parameters
  and Neutrino Masses from the Final Planck and Full-Shape BOSS Data}},
  \href{https://doi.org/10.1103/PhysRevD.101.083504}{\emph{Phys. Rev. D}
  {\bfseries 101} (2020) 083504}
  [\href{https://arxiv.org/abs/1912.08208}{{\ttfamily 1912.08208}}].

\bibitem{DAmico:2020ods}
G.~D'Amico, L.~Senatore, P.~Zhang and H.~Zheng, \emph{{The Hubble Tension in
  Light of the Full-Shape Analysis of Large-Scale Structure Data}},
  \href{https://doi.org/10.1088/1475-7516/2021/05/072}{\emph{JCAP} {\bfseries
  05} (2021) 072} [\href{https://arxiv.org/abs/2006.12420}{{\ttfamily
  2006.12420}}].

\bibitem{Zhang:2021yna}
P.~Zhang, G.~D'Amico, L.~Senatore, C.~Zhao and Y.~Cai, \emph{{BOSS Correlation
  Function analysis from the Effective Field Theory of Large-Scale Structure}},
  \href{https://doi.org/10.1088/1475-7516/2022/02/036}{\emph{JCAP} {\bfseries
  02} (2022) 036} [\href{https://arxiv.org/abs/2110.07539}{{\ttfamily
  2110.07539}}].

\bibitem{Ivanov:2021fbu}
M.~M. Ivanov, O.~H.~E. Philcox, M.~Simonovi\'c, M.~Zaldarriaga, T.~Nischimichi
  and M.~Takada, \emph{{Cosmological constraints without nonlinear
  redshift-space distortions}},
  \href{https://doi.org/10.1103/PhysRevD.105.043531}{\emph{Phys. Rev. D}
  {\bfseries 105} (2022) 043531}
  [\href{https://arxiv.org/abs/2110.00006}{{\ttfamily 2110.00006}}].

\bibitem{DAmico:2021ymi}
G.~D'Amico, L.~Senatore, P.~Zhang and T.~Nishimichi, \emph{{Taming
  redshift-space distortion effects in the EFTofLSS and its application to
  data}},  \href{https://arxiv.org/abs/2110.00016}{{\ttfamily 2110.00016}}.

\bibitem{Hamilton:2000du}
A.~J.~S. Hamilton and M.~Tegmark, \emph{{The Real space power spectrum of the
  PSCz survey from 0.01 to 300 h Mpc**-1}},
  \href{https://doi.org/10.1046/j.1365-8711.2002.05033.x}{\emph{Mon. Not. Roy.
  Astron. Soc.} {\bfseries 330} (2002) 506}
  [\href{https://arxiv.org/abs/astro-ph/0008392}{{\ttfamily
  astro-ph/0008392}}].

\bibitem{Tegmark:2001jh}
M.~Tegmark, A.~J.~S. Hamilton and Y.-Z. Xu, \emph{{The Power spectrum of
  galaxies in the 2dF 100k redshift survey}},
  \href{https://doi.org/10.1046/j.1365-8711.2002.05622.x}{\emph{Mon. Not. Roy.
  Astron. Soc.} {\bfseries 335} (2002) 887}
  [\href{https://arxiv.org/abs/astro-ph/0111575}{{\ttfamily
  astro-ph/0111575}}].

\bibitem{SDSS:2003tbn}
{\scshape SDSS} collaboration, \emph{{The 3-D power spectrum of galaxies from
  the SDSS}}, \href{https://doi.org/10.1086/382125}{\emph{Astrophys. J.}
  {\bfseries 606} (2004) 702}
  [\href{https://arxiv.org/abs/astro-ph/0310725}{{\ttfamily
  astro-ph/0310725}}].

\bibitem{Scoccimarro:2004tg}
R.~Scoccimarro, \emph{{Redshift-space distortions, pairwise velocities and
  nonlinearities}},
  \href{https://doi.org/10.1103/PhysRevD.70.083007}{\emph{Phys. Rev. D}
  {\bfseries 70} (2004) 083007}
  [\href{https://arxiv.org/abs/astro-ph/0407214}{{\ttfamily
  astro-ph/0407214}}].

\bibitem{Chen:2022jzq}
S.-F. Chen, M.~White, J.~DeRose and N.~Kokron, \emph{{Cosmological analysis of
  three-dimensional BOSS galaxy clustering and Planck CMB lensing cross
  correlations via Lagrangian perturbation theory}},
  \href{https://doi.org/10.1088/1475-7516/2022/07/041}{\emph{JCAP} {\bfseries
  07} (2022) 041} [\href{https://arxiv.org/abs/2204.10392}{{\ttfamily
  2204.10392}}].

\bibitem{Philcox:2021kcw}
O.~H.~E. Philcox and M.~M. Ivanov, \emph{{BOSS DR12 full-shape cosmology:
  \ensuremath{\Lambda}CDM constraints from the large-scale galaxy power
  spectrum and bispectrum monopole}},
  \href{https://doi.org/10.1103/PhysRevD.105.043517}{\emph{Phys. Rev. D}
  {\bfseries 105} (2022) 043517}
  [\href{https://arxiv.org/abs/2112.04515}{{\ttfamily 2112.04515}}].

\bibitem{Ivanov:2023qzb}
M.~M. Ivanov, O.~H.~E. Philcox, G.~Cabass, T.~Nishimichi, M.~Simonovi\'c and
  M.~Zaldarriaga, \emph{{Cosmology with the Galaxy Bispectrum Multipoles:
  Optimal Estimation and Application to BOSS Data}},
  \href{https://arxiv.org/abs/2302.04414}{{\ttfamily 2302.04414}}.

\bibitem{DAmico:2022osl}
G.~D'Amico, Y.~Donath, M.~Lewandowski, L.~Senatore and P.~Zhang, \emph{{The
  BOSS bispectrum analysis at one loop from the Effective Field Theory of
  Large-Scale Structure}},  \href{https://arxiv.org/abs/2206.08327}{{\ttfamily
  2206.08327}}.

\bibitem{Ivanov:2021zmi}
M.~M. Ivanov, \emph{{Cosmological constraints from the power spectrum of eBOSS
  emission line galaxies}},
  \href{https://doi.org/10.1103/PhysRevD.104.103514}{\emph{Phys. Rev. D}
  {\bfseries 104} (2021) 103514}
  [\href{https://arxiv.org/abs/2106.12580}{{\ttfamily 2106.12580}}].

\bibitem{Semenaite:2021aen}
A.~Semenaite et~al., \emph{{Cosmological implications of the full shape of
  anisotropic clustering measurements in BOSS and eBOSS}},
  \href{https://doi.org/10.1093/mnras/stac829}{\emph{Mon. Not. Roy. Astron.
  Soc.} {\bfseries 512} (2022) 5657}
  [\href{https://arxiv.org/abs/2111.03156}{{\ttfamily 2111.03156}}].

\bibitem{Semenaite:2022unt}
A.~Semenaite, A.~G. S\'anchez, A.~Pezzotta, J.~Hou, A.~Eggemeier, M.~Crocce
  et~al., \emph{{Beyond $\Lambda$CDM constraints from the full shape clustering
  measurements from BOSS and eBOSS}},
  \href{https://arxiv.org/abs/2210.07304}{{\ttfamily 2210.07304}}.

\bibitem{Chudaykin:2022nru}
A.~Chudaykin and M.~M. Ivanov, \emph{{Cosmological constraints from the power
  spectrum of eBOSS quasars}},
  \href{https://arxiv.org/abs/2210.17044}{{\ttfamily 2210.17044}}.

\bibitem{Simon:2022csv}
T.~Simon, P.~Zhang and V.~Poulin, \emph{{Cosmological inference from the
  EFTofLSS: the eBOSS QSO full-shape analysis}},
  \href{https://arxiv.org/abs/2210.14931}{{\ttfamily 2210.14931}}.

\bibitem{Neveux:2022tuk}
R.~Neveux et~al., \emph{{Combined full shape analysis of BOSS galaxies and
  eBOSS quasars using an iterative emulator}},
  \href{https://doi.org/10.1093/mnras/stac2114}{\emph{Mon. Not. Roy. Astron.
  Soc.} {\bfseries 516} (2022) 1910}
  [\href{https://arxiv.org/abs/2201.04679}{{\ttfamily 2201.04679}}].

\bibitem{Brieden:2022heh}
S.~Brieden, H.~Gil-Mar\'\i{}n and L.~Verde, \emph{{A tale of two (or more)
  $h$'s}},  \href{https://arxiv.org/abs/2212.04522}{{\ttfamily 2212.04522}}.

\bibitem{Bahr-Kalus:2023ebd}
B.~Bahr-Kalus, D.~Parkinson and E.-M. Mueller, \emph{{Measurement of the
  matter-radiation equality scale using the extended Baryon Oscillation
  Spectroscopic Survey Quasar Sample}},
  \href{https://arxiv.org/abs/2302.07484}{{\ttfamily 2302.07484}}.

\bibitem{Smith:2022iax}
T.~L. Smith, V.~Poulin and T.~Simon, \emph{{Assessing the robustness of sound
  horizon-free determinations of the Hubble constant}},
  \href{https://arxiv.org/abs/2208.12992}{{\ttfamily 2208.12992}}.

\bibitem{Chudaykin:2020ghx}
A.~Chudaykin, K.~Dolgikh and M.~M. Ivanov, \emph{{Constraints on the curvature
  of the Universe and dynamical dark energy from the Full-shape and BAO data}},
  \href{https://doi.org/10.1103/PhysRevD.103.023507}{\emph{Phys. Rev. D}
  {\bfseries 103} (2021) 023507}
  [\href{https://arxiv.org/abs/2009.10106}{{\ttfamily 2009.10106}}].

\bibitem{Chudaykin:2022rnl}
A.~Chudaykin, D.~Gorbunov and N.~Nedelko, \emph{{Exploring $\Lambda$CDM
  extensions with SPT-3G and Planck data: 4$\sigma$ evidence for neutrino
  masses, full resolution of the Hubble crisis by dark energy with phantom
  crossing, and all that}},  \href{https://arxiv.org/abs/2203.03666}{{\ttfamily
  2203.03666}}.

\bibitem{Vagnozzi:2020rcz}
S.~Vagnozzi, E.~Di~Valentino, S.~Gariazzo, A.~Melchiorri, O.~Mena and J.~Silk,
  \emph{{The galaxy power spectrum take on spatial curvature and cosmic
  concordance}}, \href{https://doi.org/10.1016/j.dark.2021.100851}{\emph{Phys.
  Dark Univ.} {\bfseries 33} (2021) 100851}
  [\href{https://arxiv.org/abs/2010.02230}{{\ttfamily 2010.02230}}].

\bibitem{Kumar:2022vee}
S.~Kumar, R.~C. Nunes and P.~Yadav, \emph{{Updating non-standard neutrinos
  properties with Planck-CMB data and full-shape analysis of BOSS and eBOSS
  galaxies}}, \href{https://doi.org/10.1088/1475-7516/2022/09/060}{\emph{JCAP}
  {\bfseries 09} (2022) 060}
  [\href{https://arxiv.org/abs/2205.04292}{{\ttfamily 2205.04292}}].

\bibitem{Reeves:2022aoi}
A.~Reeves, L.~Herold, S.~Vagnozzi, B.~D. Sherwin and E.~G.~M. Ferreira,
  \emph{{Restoring cosmological concordance with early dark energy and massive
  neutrinos?}}, \href{https://doi.org/10.1093/mnras/stad317}{\emph{Mon. Not.
  Roy. Astron. Soc.} {\bfseries 520} (2023) 3688}
  [\href{https://arxiv.org/abs/2207.01501}{{\ttfamily 2207.01501}}].

\bibitem{DAmico:2020kxu}
G.~D'Amico, L.~Senatore and P.~Zhang, \emph{{Limits on $w$CDM from the EFTofLSS
  with the PyBird code}},
  \href{https://doi.org/10.1088/1475-7516/2021/01/006}{\emph{JCAP} {\bfseries
  01} (2021) 006} [\href{https://arxiv.org/abs/2003.07956}{{\ttfamily
  2003.07956}}].

\bibitem{DAmico:2020tty}
G.~D'Amico, Y.~Donath, L.~Senatore and P.~Zhang, \emph{{Limits on Clustering
  and Smooth Quintessence from the EFTofLSS}},
  \href{https://arxiv.org/abs/2012.07554}{{\ttfamily 2012.07554}}.

\bibitem{Ivanov:2020ril}
M.~M. Ivanov, E.~McDonough, J.~C. Hill, M.~Simonovi\'c, M.~W. Toomey,
  S.~Alexander et~al., \emph{{Constraining Early Dark Energy with Large-Scale
  Structure}}, \href{https://doi.org/10.1103/PhysRevD.102.103502}{\emph{Phys.
  Rev. D} {\bfseries 102} (2020) 103502}
  [\href{https://arxiv.org/abs/2006.11235}{{\ttfamily 2006.11235}}].

\bibitem{Simon:2022adh}
T.~Simon, P.~Zhang, V.~Poulin and T.~L. Smith, \emph{{Updated constraints from
  the effective field theory analysis of BOSS power spectrum on Early Dark
  Energy}},  \href{https://arxiv.org/abs/2208.05930}{{\ttfamily 2208.05930}}.

\bibitem{Euclid:2019clj}
{\scshape Euclid} collaboration, \emph{{Euclid preparation: VII. Forecast
  validation for Euclid cosmological probes}},
  \href{https://doi.org/10.1051/0004-6361/202038071}{\emph{Astron. Astrophys.}
  {\bfseries 642} (2020) A191}
  [\href{https://arxiv.org/abs/1910.09273}{{\ttfamily 1910.09273}}].

\bibitem{Euclid:2021qvm}
{\scshape Euclid} collaboration, \emph{{Euclid preparation - XV. Forecasting
  cosmological constraints for the Euclid and CMB joint analysis}},
  \href{https://doi.org/10.1051/0004-6361/202141556}{\emph{Astron. Astrophys.}
  {\bfseries 657} (2022) A91}
  [\href{https://arxiv.org/abs/2106.08346}{{\ttfamily 2106.08346}}].

\bibitem{Chudaykin:2019ock}
A.~Chudaykin and M.~M. Ivanov, \emph{{Measuring neutrino masses with
  large-scale structure: Euclid forecast with controlled theoretical error}},
  \href{https://doi.org/10.1088/1475-7516/2019/11/034}{\emph{JCAP} {\bfseries
  11} (2019) 034} [\href{https://arxiv.org/abs/1907.06666}{{\ttfamily
  1907.06666}}].

\bibitem{Zhang:2019dyq}
J.-F. Zhang, L.-Y. Gao, D.-Z. He and X.~Zhang, \emph{{Improving cosmological
  parameter estimation with the future 21 cm observation from SKA}},
  \href{https://doi.org/10.1016/j.physletb.2019.135064}{\emph{Phys. Lett. B}
  {\bfseries 799} (2019) 135064}
  [\href{https://arxiv.org/abs/1908.03732}{{\ttfamily 1908.03732}}].

\bibitem{Sprenger:2018tdb}
T.~Sprenger, M.~Archidiacono, T.~Brinckmann, S.~Clesse and J.~Lesgourgues,
  \emph{{Cosmology in the era of Euclid and the Square Kilometre Array}},
  \href{https://doi.org/10.1088/1475-7516/2019/02/047}{\emph{JCAP} {\bfseries
  02} (2019) 047} [\href{https://arxiv.org/abs/1801.08331}{{\ttfamily
  1801.08331}}].

\bibitem{Sailer:2021yzm}
N.~Sailer, E.~Castorina, S.~Ferraro and M.~White, \emph{{Cosmology at high
  redshift \textemdash{} a probe of fundamental physics}},
  \href{https://doi.org/10.1088/1475-7516/2021/12/049}{\emph{JCAP} {\bfseries
  12} (2021) 049} [\href{https://arxiv.org/abs/2106.09713}{{\ttfamily
  2106.09713}}].

\bibitem{Ferraro:2022cmj}
S.~Ferraro, N.~Sailer, A.~Slosar and M.~White, \emph{{Snowmass2021 Cosmic
  Frontier White Paper: Cosmology and Fundamental Physics from the
  three-dimensional Large Scale Structure}},
  \href{https://arxiv.org/abs/2203.07506}{{\ttfamily 2203.07506}}.

\end{thebibliography}\endgroup

\end{document}